\documentstyle[aps,psfig,prb,multicol]{revtex}
\begin{document}
\draft
\title{Mobility Edge in Aperiodic Kronig-Penney Potentials with
Correlated Disorder: Perturbative Approach}
\author{F.M. Izrailev,$^a$ A.A. Krokhin,$^{a,b}$ and S.E. Ulloa$^b$}
\address{$^a$Instituto de F\'\i sica, Universidad Aut\'onoma
de Puebla, Apartado Postal J-48, Pue., 72570, M\'exico \\
$^b$Department of Physics and Astronomy and Condensed Matter and
Surface Sciences Program\\ Ohio University, Athens, OH 45701-2979}
\date{\today}
\maketitle

\begin{abstract}
It is shown that a non-periodic Kronig-Penney model exhibits
mobility edges if the positions of the scatterers are correlated
at long distances. An analytical expression for the
energy-dependent localization length is derived for weak disorder
in terms of the real-space correlators defining the structural
disorder in these systems. We also present an algorithm to
construct a non-periodic but correlated sequence exhibiting
desired mobility edges.  This result could be used to construct
window filters in electronic, acoustic, or photonic non-periodic
structures.

\end{abstract}
\begin{multicols}{2}

 The Kronig-Penney model has been widely used to explore the
characteristics of electrons in a periodic potential, as this
model provides one with perhaps the simplest instance of Bloch
states. This model is also used systematically to provide
estimates of the bandwidths in semiconductor superlattices with
high reliability. \cite{Bastard-book} The Kronig-Penney model and
its relation with superlattices has also been used in recent
times to provide an implementation of the physics of random and
quasi-periodic systems, \cite{DS-Prange} and interesting
experiments have been reported on arrangements such as the
well-known Fibonacci sequences. \cite{Merlin} It is because of
its importance and wide applicability that we focus our attention
on the Kronig-Penney model. We will demonstrate that the
aperiodic model with constant scattering potential but random
spacings yields mobility edges if the disorder has long-range
correlations, in sharp contrast to the situation for white noise
potentials.

The Kronig-Penney model in this study is given by a 1D chain of
delta-function scatterers with amplitude $U_n$ and centered at
points $z_n$. The Schr\"odinger equation for a particle moving in
this random potential has the form,
\begin{equation}
(\hbar ^2/2m )\psi ^{\prime \prime }(z)+E\psi
(z)=\sum_{n=-\infty }^\infty U_n\psi (z_n)\delta (z-z_n)\, .
\label{Schr}
\end{equation}
This is equivalent to the discrete equation,
\begin{equation}
\begin{array}{cc}
& \sin \mu _{n-1} \psi _{n+1}  + \sin \mu _{n}\psi _{n-1} = \nonumber \\
& \left[ \sin (\mu _{n}+   \mu _{n-1})  +
 (U _{n}/q) \sin \mu _{n-1} \sin \mu _{n} \right] \psi _{n},
\label{discr}
\end{array}
\end{equation}
where $\psi _{n}\equiv \psi (z_{n})$, $\mu
_{n}=q(z_{n+1}-z_{n})$, $q=\sqrt{E}$, and the energy is measured
in units where $\hbar ^2/2m=1$. The linear relation (\ref{discr})
between $\psi_{n-1}$, $\psi_{n}$, and $\psi_{n+1}$,  can be
easily arrived at by integrating Eq.\ (\ref{Schr}) in the
vicinity of sites ${n-1}$, $n$, and $n+1$, and substituting the
amplitudes for the various constants in the piece-wise
zero-potential regions between the scatterers. If the site
potential is different from a delta function, a linear relation
similar to Eq.\ (\ref{discr}) can be obtained using a general
method described in Ref.\ \onlinecite{SMDA94}. Therefore, Eq.\
(\ref{discr}) can be considered as a rather generic relation for
1D chains with potential scatterers. We focus here on a sequence
of scatterers of {\em equal} amplitude, $U_n \equiv U$, but with
varying positions, i.e. the case of `structural' or `positional'
disorder. An obvious experimental realization of this model is a
semiconductor superlattice with fluctuating period.

The case of {\em compositional} disorder, i.e., a periodic
arrangement of sites, $z_n=n$, with random amplitude $U_n$, has
been studied intensively during the last decade. The importance of
short-range correlations was first explored recently, \cite{F89}
and a `random dimer' model was studied as a specific example.
\cite{dun90} Using this model, it was shown that short-range
correlations in the infinite random sequence $\left \{U_n \right
\}$ give rise to a {\it discrete} number of delocalized states as
well as to some anomalies in transport properties. \cite{Bov92}
The presence of such anomalies has been recently observed in
experiments with GaAs-AlGaAs random-dimer superlattices.
\cite{Bel99} Moreover, the non-trivial role of correlations in
the formation of mobility edges has been pointed out by the study
of localization in pseudorandom and incommensurate potentials.
\cite{Sarma88}  In contrast, only relatively recently the role
of long-range correlations in random potentials has received special
attention.   The interplay of long-range correlations and
disorder has been shown to lead to the existence of a {\it
continuum} of extended states in the energy spectrum and to the
appearance of mobility edges.\cite{mou98,izr99} These edges have
been shown to exist in experiments of microwave transmission in a
single-mode waveguide with a random array of correlated
scatterers,\cite{Koh00}  and their possible relevance for
metal-insulator transitions in 2D electron systems has been
recently explored.\cite{FV99}

{\em Localization length for weak disorder}. To calculate the
localization length $l(E)$ in the case of {\em structural
disorder} we use a Hamiltonian approach. \cite{izr95} In this
scheme, the discrete Schr\"odinger equation is replaced by a
classical Hamiltonian map for coordinate $x_n$ and momentum
$p_n$. For the case of Eq.\ (\ref{discr}), conjugate variables
are introduced as, $\,x_{n}=\psi _{n}$ and $p_{n}=(x_n \cos
\mu_{n-1} - x_{n-1})/ \sin \mu_{n-1}$.  Correspondingly, the
discrete-time evolution of $x_n$ and $p_n$ is obtained from
\begin{equation}
\begin{array}{cc}
 p_{n+1}=(p_{n}\,+A_{n}\,x_{n})\cos \mu _{n}\,\,-\,\,x_{n} \sin
 \mu _{n}, &
\\
 x_{n+1}=(p_{n}\,+A_{n}\,x_{n})\sin \mu _{n}\,\,+\,\,x_{n} \cos
 \mu_{n} \, .
\end{array}
\label{map}
\end{equation}
This map describes the behavior of a linear rotator subjected to
{\it non-periodic} delta-kicks with amplitude $A_n=U_n/q$. Free
rotation between kicks corresponds to free propagation between
scatterers and each kick corresponds to scattering at each
$\delta$-function potential. It is easy to check that the first
equation in (\ref{map}) is equivalent to (\ref{discr}), while the
second is reduced to an identity after $p_n$ substitution.

In what follows we consider the case of weak disorder, assuming
that the deviation of the scatterers from their positions in a
periodic lattice is small, $| \delta _{n}| = q | z_n - n | \ll 1$.
We can then expand trigonometric functions in (\ref{map}) and up
to second order in $\lambda _{n}=q(\delta _{n+1}-\delta _{n})$,
obtain the approximate map for constant kick amplitude $A=U/q$,
\begin{equation}
\begin{array}{cc}
 p_{n+1} =  \cos q \left[ p_n(1-\frac{\lambda _n^2}{2})  +
x_n ( A-\lambda _n - A \frac{\lambda _n^2}{2})\right]\\
-\sin q \left[p_n\lambda _n + x_n(1+ A\lambda _n - \frac{\lambda
_n^2}{2})
\right], \\
  x_{n+1} =  \cos q\left[p_n\lambda_n + x_n(1 + A\lambda _n
-\frac{\lambda _n^2}{2})\right] \\
+ \sin q \left[p_n(1 -\frac{\lambda _n^2}{2}) + x_n(A - \lambda
_n - A \frac{\lambda _n^2}{2}) \right].  \label{map1}
\end{array}
\end{equation}
Note that since $\lambda _n \propto q =\sqrt E$, this expansion
is valid only for low energies.

In order to extract the effect that comes only from the positional
non-periodicity, it is convenient to eliminate the mean field
associated with the constant amplitude $U$. This can be done by a
canonical transformation of variables,
\begin{equation}
\begin{array}{cc}
p_{n}=\alpha ^{-1}\cos \phi \, P_{n}\,-\alpha \sin \phi \,
 X_{n}\,, &  \\
x_{n}=\alpha ^{-1}\sin \phi \, P_{n}\,+\alpha \cos \phi \,
 X_{n}\,. &
\end{array}
\label{trans}
\end{equation}
The parameters of this transformation ($\alpha $ and $\phi $) are
obtained from the condition that to zeroth order in $\lambda _n$
the dynamical map for $(P_n,X_n)$ be a simple rotation without
kicks, i.e., $P_{n+1}^{(0)}=P_{n}^{(0)}\cos \gamma
-X_{n}^{(0)}\sin \gamma $, and $X_{n+1}^{(0)}=P_{n}^{(0)}\sin
\gamma +X_{n}^{(0)}\cos \gamma$. Applying this condition to Eqs.\
(\ref{trans}) and (\ref{map1}) we get after some algebra:
\begin{equation}
\begin{array}{cc}
\label{phi}
 \phi = \frac{1}{2} q\,,\,\,\, \cos \gamma =\cos q
 +(A/2q)\sin q\,, \\
 \alpha ^{4}= 1+ 2A/ \left[2\sin q- A (1+\cos q)\right] .
\end{array}
\end{equation}
It is clear that $\gamma$ plays the role of the Bloch number in
the periodic Kronig-Penney model. \cite{Bastard-book}  We can now
rewrite the map (\ref{map1}) in terms of variables $(P_n,X_n)$
and angle $\gamma $,
\begin{equation}
\begin{array}{cc}
P_{n+1} =  \left( 1-\lambda _{n}^{2}/2\right) (P_{n}\cos
\gamma -X_{n}\sin \gamma ) & \\ - \lambda _{n}\alpha
^{2}(P_{n}\sin \gamma +X_{n}\cos
\gamma ), \\
X_{n+1} = \left( 1-\lambda _{n}^{2}/2\right) (P_{n}\sin
\gamma +X_{n}\cos \gamma ) & \\ + \lambda _{n}\alpha
^{-2}(P_{n}\cos \gamma -X_{n}\sin \gamma ).
\end{array}
\label{newmap1}
\end{equation}

Following our previous approach,\cite{izr99} it is convenient to
introduce action-angle variables for map (\ref{newmap1}),
\begin{equation}
P_{n}=R_{n}\sin \theta _{n},\,\,\,\, X_{n}=R_{n}\cos \theta _{n}
\, . \label{aavar}
\end{equation}
The inverse localization length for the original quantum model
Eq.\ (\ref{Schr}) (or equivalently, the Lyapunov exponent for the
dynamical map in (\ref{map})), can be expressed in terms of the
ratio $R_{n+1}/R_n$ (see details in Ref.\ \onlinecite{izr95}),
\begin{eqnarray}
\label{length}
 l^{-1}(E)=\left\langle {\ln
\frac{\psi_{n+1}}{\psi_n}}\right\rangle &=&
 \left\langle {\ln \frac{X_{n+1}}{X_n}}\right\rangle
 = \frac{1}{2}
\left\langle {\ln \frac{R_{n+1}^2}{R_n^2}}\right\rangle \, .
\end{eqnarray}
Here, angular brackets stand for the average over sites (kicks),
so that, $\left\langle {...}\right\rangle = \lim _{N\rightarrow
\infty }\,\frac 1N\sum_{n=1}^N \left( ...\right)$. Using the map
(\ref{newmap1}) we can calculate the ratio $R_{n+1}/R_n$ as
\begin{equation}
\begin{array}{cc}
\label{r/r}
\left( R_{n+1}/R_{n}\right) ^{2} =
1-U \lambda _{n} \sin \left[ 2\left( \theta
_{n}-\gamma \right)
\right ] / (q\sin \gamma) \\
+ \lambda _{n}^{2} \left[
 \alpha ^{-4} \sin ^{2}(\theta _n - \gamma ) + \alpha ^{4} \cos ^2 (\theta _n - \gamma) -1 \right]
\end{array}
\end{equation}
The fact that the ratio
$R_{n+1}/R_n$ is close to unity for small $\lambda _n$ is what
motivated switching from the variables ($p_n,x_n$) to ($P_n,X_n$)
using (\ref{trans}). The logarithm in (\ref{length}) can be
expanded as $\ln (1+x)\approx x-x^2/2$, and, up to second order
in $\lambda _n$, the average is performed over the unperturbed
motion given by the $(P_{n}^{(0)},X_n^{(0)})$ variables.  Since
this motion is a {\it free} rotation (in the old variables it is
a rotation with {\it periodic kicks}), the angle variable $\theta
_n$ is clearly distributed uniformly within the interval
$[0,2\pi]$. One then obtains that $\left\langle \sin ^2 \theta _n
\right\rangle = \left\langle{\cos^2 \theta _n } \right\rangle =
1/2$, and
\begin{equation}
\label{length1} l^{-1}(E)=
\frac{\left\langle{\lambda_n^2}\right\rangle U ^2} {8q^2 \sin^2
\gamma }  - \frac{U }{q \sin \gamma} \left\langle {\lambda
_n\sin[2(\theta_n - \gamma)]} \right\rangle\, .
\end{equation}
The first term in (\ref{length1}) gives the inverse localization
length in an uncorrelated random potential. In this Born
approximation, it is proportional to the variance $\left\langle
\lambda_n^2 \right\rangle$ and to the squared amplitude of the
scattering potential, $U^2$. Since $\left\langle{\lambda_n^2}
\right\rangle\propto q^2$, the factor $q^2$ disappears and the
only energy dependence is due to the factor $\sin^2\gamma $ in
the denominator. At the edges of the allowed zones $\sin \gamma =
0$, and here the localization length $l(E)$ approaches zero.  A
similar enhancement of localization in the vicinity of the band
edges \cite{John,Marad} has stimulated the study of photonic-band-gap
materials in the last decade.

The second term in (\ref{length1}) describes the contribution of
correlations in the scattering potential. To  calculate explicitly
the correlator $\left\langle{\lambda _n\sin[2(\theta_n -
\gamma)]} \right\rangle $, one needs the recursion relation for
the angle variable $\theta_n$. Since this correlator already
contains a factor $\lambda_n$, only linear terms in the recursion
relation are needed from Eqs.\ (\ref{newmap1}) and (\ref{aavar}),
so that $\theta _{n} =  \theta _{n-1}-\gamma -\lambda _{n-1}
\left [ \alpha ^2 - U \sin^2(\theta _{n-1}
- \gamma )/(q \sin \gamma ) \right ]$. The correlator $\left\langle{\lambda _n
\sin[2(\theta_n -\gamma)]} \right\rangle $ can be written as a
Fourier series in Bloch number $\gamma$, where the dimensionless
correlators $\xi(k)$ are the Fourier coefficients,\cite{izr99}
\begin{equation}
\label{corr3} \left\langle{\lambda _n\sin[2(\theta_n - \gamma)]}
\right\rangle  = - \frac{U q \Delta ^2}{2 \sin\gamma }
\sum_{n=1}^{\infty }\xi(n)\cos(2 \gamma n)\, .
\end{equation}
Here $\Delta^2= \left\langle{\Delta _n^2} \right\rangle$, with
$\xi(k)=\left\langle{\Delta _{n+k}\Delta_{n}} \right\rangle
/\Delta^2$, and  $\Delta_n=\delta_{n+1}-\delta_n $. Note that the
localization length is determined by the statistical properties
of the sequence of {\em relative} displacements $\Delta_n$, and
not by the displacements $\delta_n$ themselves.

Substituting the correlator (\ref{corr3}) into (\ref{length1}) we
get the final result for the inverse localization length,
\begin{equation}
\label{length2}
\frac{1}{l(E)}= \frac{U^2 \Delta ^2}{8 \sin^2 \gamma}\varphi
(\gamma ), \,\, \varphi (\gamma ) =  1+
2 \sum_{n=1}^{\infty }\xi(n)\cos(2 \gamma n) .
\end{equation}
This formula has the same structure as that obtained for a
tight-binding (and the corresponding Kronig-Penney) model with
random amplitudes $U_n$, but equidistant sites ($z_n=n$).
\cite{izr99,Koh00}  These three different models have a different
dependence on energy via the factors $q^2$ and $\sin^2 \gamma$.
The present case exhibits the weakest dependence of the
localization length on energy within the allowed zone since a
factor $q^2=E$ appears in other cases but not here. This property
should be favorable for the experimental observation of mobility
edges in superlattices with positional disorder.

{\em Correlations and mobility edge}. If the sequence of random
displacements $\Delta _n$ is uncorrelated, $\xi (k)=0$, the
localization length is given by the first term in
(\ref{length2}). In the opposite limit of a completely correlated
sequence, $\xi(k)=$const., the displacements are independent on
the site number, $\delta_n=\delta_0$, giving a regular sequence
with period $1+\delta_0$, and extended states, $l^{-1}=0$, for
all energies. A smooth transition between these two limits can be
described by an exponential function, $\xi(k) = \exp(-k/k_0)$,
where $k_0$ is a correlation radius. Substituting this form into
(\ref{length2}) one obtains
\begin{equation}
\label{locexp} l^{-1}(E)={\frac{U {^2}\Delta^2}{8 \,{\sin
^2\gamma}}\,}{\frac{{ \sinh {(1/k_0)} }}{{\cosh (1/k_0) -\cos
(2\gamma )}}}\,.
\end{equation}
For any finite $k_0$ {\em all} states are localized.  {\it Only}
in a periodic lattice ($k_0=\infty$) does the inverse
localization length (\ref{locexp}) vanish and the states become
delocalized.  This localization-delocalization transition occurs
simultaneously for all energies, and a mobility edge does not
appear in the spectrum. This conclusion is valid for arbitrary
amplitude of the scattering potential $U$.\cite{CPV89}  A
discrete number of delocalized states can in fact appear if the
binary correlator $\xi(k)$ oscillates with exponentially decaying
amplitude.\cite{DR79} This can be obtained from (\ref{locexp}),
if the correlation radius $k_0$ is allowed to take complex
values. On the other hand, a mobility edge may appear if
correlations decay not exponentially but according to a power
law.\cite{mou98,izr99,Koh00}  We show numerically below that mobility
edges exist if, e.g., $\xi(k)\propto 1/k$.

{\em Designed mobility edges}. In a real (or numerical) experiment
one needs to know explicitly the displacements $\delta_n$ which
provide a desirable dependence $l(E)$. This leads us to the
`inverse problem' in the theory of localization.  The solution
would give us in general a random scattering potential $\{U_n,
\delta _n\}$ through the dependence $l(E)$. In the present case of
positional disorder, $U_n=U$, we need to calculate only the
relative displacements. One can explicitly evaluate the
correlator $\xi(k)$ as the Fourier coefficient,
\begin{equation}
\xi (k)=\frac 2\pi \int_0^{\pi /2}\varphi (\gamma )\cos (2k\gamma
)d\gamma \,, \label{corr}
\end{equation}
where the energy dependence of
$\varphi (\gamma )=8 \sin^2\gamma /(U^2 \Delta^2 l(E))$,
is assumed to be known.  The energy $E=q^2$ is expressed
through $\gamma$ via the dispersion relation in (\ref{phi}). It
is easy to check that the binary correlator of a sequence
$(\Delta _n /\Delta) =\sum_{k=-\infty }^\infty \beta (k)\,Z_{n+k}
\, $, coincides with $ \xi(k)$ if $Z_k$ are random numbers with
zero mean and unit variance, and where the function $\beta(k)$ is
given by \cite{izr99,Koh00,ThanksSok}
\begin{equation}
\beta (k)=\frac 2\pi \int_0^{\pi /2}\sqrt{\varphi (\gamma )}\cos
(2k \gamma )d\gamma \,.  \label{beta}
\end{equation}
Once the relative displacements $\Delta_n$ are known, a sequence
of absolute displacements can be easily calculated, setting
$\delta_0=0$, for example, and then $\delta_1 = \Delta_0,$
$\delta_2 = \Delta_0+\Delta_1,$ $\ldots  $\,, $\delta_n  =
\sum_{k=0}^{n-1} \Delta_k$.  This procedure allows the
calculation of displacements $\delta_n$ for {\it any} energy
dependence of the localization length $l(E)$, including
situations with mobility edges. Appropriately correlated elements
may be used for fabrication of effective filters of electrical or
optical signals, even if the system is {\em not} periodic.
\cite{class-waves} The bandwidth of a filter can be made
arbitrarily wide or narrow depending on the {\em statistical
properties} of the random sequence used (via the function
$\beta(k))$.

{\em Numerical examples}.  In order to examine our predictions, we
construct explicit random sequences $\{\delta _n \}$ for which
the function $\varphi (\gamma )$ in (\ref{length2}) has four
mobility edges at $\gamma _i = 0.2\pi i,\,\, i=1, \dots, 4$. The
positions of the mobility edges are chosen within the interval $0
<\gamma <\pi$, symmetrically about $\gamma =\pi/2$. Via relation
(\ref{phi}), the position of the mobility edges on the axis
$q/\pi$ are given by 0.326, 0.478, 0.649, and 0.850, for the mean
field amplitude $U=0.7$. In the ideal lattice with the same
strength of the potential, the first allowed band lies between
$q_l/\pi \approx 0.26$ ($\gamma =0$) and $q_r/\pi \approx 1$ ($\gamma
=\pi$). Numerical data are given for two complimentary situations
$\varphi _{1,2}(\gamma )$:  $\varphi _1$ vanishes (i.e.
$l(E)=\infty $ and states are delocalized) in the region $\gamma
\in (0, \gamma_1) \cup (\gamma_2, \gamma_3) \cup (\gamma_4, \pi)$;
while $\varphi _2$ vanishes in the complementary region $\gamma \in
(\gamma_1, \gamma_2) \cup (\gamma_3, \gamma_4)$.
Outside these regions, the function $\varphi _{1,2}(\gamma )$ is
a constant defined by the normalization condition $\xi
_{1,2}(0)=1$. From Eq.\ (\ref{corr}), we get that because of the
presence of sharp mobility edges the correlators (Fourier
components of a discontinuous function) decay slowly:
\begin{eqnarray}
\label{xi_1,2}
& \xi _1(k)=-1.5 \, \xi _2(k) = \\ \nonumber
& (5/2\pi k)\left[ \sin(0.8\pi k) - \sin (0.4\pi k) \right].
\end{eqnarray}
\newpage
\vspace{-1.5cm}
\begin{figure}[p]
\hspace{-2.3cm}
\psfig{file=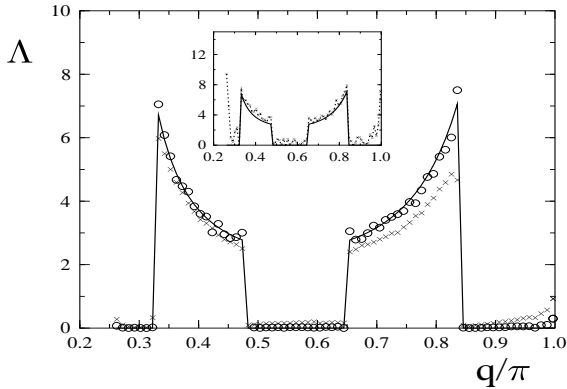,width=8cm,height=8cm,angle=-90}
\vspace{-0.5cm}
\narrowtext
\caption{Inverse localization length for band
with four mobility edges, three regions of extended states, and
different amplitude of disorder: circles ($\circ$), $\Delta =
0.05$; crosses ($\times$), $\Delta = 0.15$.  Solid lines show
Eq.\ ({\protect \ref{length2}}) for $\Delta =0.05$.  Inset shows
$N=10^3$ sites averaged over only 100 samplings.}
\end{figure}
\vspace{-0.7cm}
\begin{figure}[p]
\hspace{-2.3cm}
\psfig{file=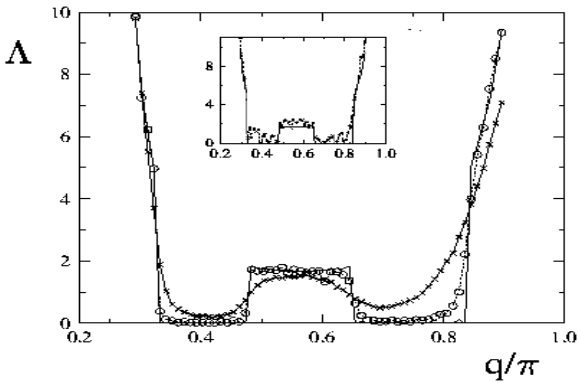,width=15.13cm,height=9.1cm,angle=-90}
\vspace{-2.0cm}
\caption{Results for complementary system to Fig.\ 1, with four
mobility edges but only two regions of extended states.  Lines
through circles ($\Delta =0.05$) and crosses ($\Delta=0.15$) are
guides to the eye.}
\end{figure}

We show the corresponding data in Fig.\ 1 and 2 for $U=0.7$. The
analytical dependence (\ref{length2}) for the dimensionless
inverse localization length $\Lambda =8/l(E)U^2\Delta^2$ is shown
by the full lines for $\Delta =0.05\,$ in the figures. Numerical
data are obtained for a large sample size $N=10^6$, with two
amplitudes of disorder, $\Delta =0.05$ (circles) and $\Delta
=0.15$ (crosses).   The insets show results for a much shorter
sample, $N=1000$, with additional average over 100 different
realizations of disorder but the same correlations. One can see
that for small disorder, $\Delta=0.05$, Eq.(\ref{length2})
describes the numerical results very well, with only minor
deviations close to the onset of mobility $q_i$ and band edges,
$q_{l,r}$ .  Notice that Eq.(\ref{length2}) is excellent at
giving the mobility edges at the prescribed energies, which
proves its usefulness, and breadth of applicability.

This work was supported by CONACyT (Mexico) Grants No.\ 26163-E
and 28626-E, and by the US Department of Energy Grant No.\
DE-FG02-91ER45334. AAK is grateful for a Rufus Putnam Fellowship
from Ohio University.

\end{multicols}
\end{document}